\documentclass{aa}  

\usepackage{graphicx}
\usepackage{amsmath}
\usepackage{amssymb}
\usepackage{hyperref}
\usepackage{times}
\usepackage{lscape}
\usepackage{afterpage}
\usepackage{ifthen}

\usepackage{xcolor}%
\usepackage{soul}

\usepackage{txfonts}
\newcommand{\FeKa}{Fe K\ensuremath{\alpha}\xspace}
\newcommand{\FeKb}{Fe K\ensuremath{\beta}\xspace}
\newcommand{\lya}{Fe Ly\ensuremath{\alpha}\xspace}
\newcommand{\lyb}{Fe Ly\ensuremath{\beta}\xspace}
\newcommand{\logxi}{\ensuremath{{\log \xi}}\xspace}
\newcommand{\NH}{\ensuremath{N_{\mathrm{H}}}\xspace}
\newcommand{\vout}{\ensuremath{v_{\mathrm{out}}}\xspace}
\newcommand{\sigv}{\ensuremath{\sigma_{v}}\xspace}

\begin{document} 

\title{Delving into the depths of NGC 3783 with XRISM}
\subtitle{III. Birth of an ultrafast outflow during a soft flare }

\author{
Liyi Gu \inst{1,2,3,4}
\and
Keigo Fukumura \inst{5}
\and
Jelle Kaastra \inst{1,2}
\and
Megan Eckart \inst{6}
\and
Ralf Ballhausen \inst{7,8,9}
\and
Ehud Behar \inst{10,11}
\and
Camille Diez \inst{12}
\and
Matteo Guainazzi \inst{13}
\and
Timothy Kallman \inst{8}
\and
Erin Kara \inst{11}
\and
Chen Li \inst{2,1}
\and
Missagh Mehdipour \inst{14}
\and
Misaki Mizumoto \inst{15}
\and
Shoji Ogawa \inst{16}
\and
Christos Panagiotou \inst{11}
\and
Matilde Signorini \inst{13,17}
\and
Atsushi Tanimoto \inst{18}
\and
Keqin Zhao \inst{2,1}
\and
Hirofumi Noda \inst{19}
\and
Jon Miller \inst{20}
\and 
Satoshi Yamada \inst{19}
}

\institute{
SRON Space Research Organisation Netherlands, Niels Bohrweg 4, 2333 CA Leiden, the Netherlands
\and
Leiden Observatory, Leiden University, PO Box 9513, 2300 RA Leiden, The Netherlands
\and
RIKEN High Energy Astrophysics Laboratory, 2-1 Hirosawa, Wako, Saitama 351-0198, Japan
\and
Department of Physics, Tokyo University of Science, 1-3 Kagurazaka, Shinjuku-ku, Tokyo 162-8601, Japan
\and
Department of Physics and Astronomy, James Madison University, Harrisonburg, VA 22807, USA
\and
Lawrence Livermore National Laboratory, Livermore, CA 94550, USA
\and
University of Maryland College Park, Department of Astronomy, College Park, MD 20742, USA
\and
NASA Goddard Space Flight Center (GSFC), Greenbelt, MD 20771, USA
\and
Center for Research and Exploration in Space Science and Technology, NASA GSFC (CRESST II), Greenbelt, MD 20771, USA
\and
Department of Physics, Technion, Haifa 32000, Israel
\and
MIT Kavli Institute for Astrophysics and Space Research, Massachusetts Institute of Technology, Cambridge, MA 02139, USA
\and
ESA European Space Astronomy Centre (ESAC), Camino Bajo del Castillo s/n, 28692 Villanueva de la Cañada, Madrid, Spain
\and
ESA European Space Research and Technology Centre (ESTEC), Keplerlaan 1, 2201 AZ, Noordwĳk, the Netherlands
\and 
Space Telescope Science Institute, 3700 San Martin Drive, Baltimore, MD 21218, USA 
\and
Science Research Education Unit, University of Teacher Education Fukuoka, Munakata, Fukuoka 811-4192, Japan
\and
Institute of Space and Astronautical Science (ISAS), Japan Aerospace Exploration Agency (JAXA), Kanagawa 252-5210, Japan
\and
INAF - Osservatorio Astrofisico di Arcetri, Largo Enrico Fermi 5, I-50125 Florence, Italy
\and
Graduate School of Science and Engineering, Kagoshima University, Kagoshima, 890-8580, Japan
\and
Astronomical Institute, Tohoku University, 6-3 Aramakiazaaoba, Aoba-ku, Sendai, Miyagi 980-8578, Japan
\and
Department of Astronomy, University of Michigan, 1085 South University Avenue, Ann Arbor, MI 48109-1107, USA
}

\abstract{
The 2024 X-ray/UV observation campaign of NGC 3783, led by XRISM, revealed the launch of an ultrafast outflow (UFO) with a radial velocity of 0.19$c$ (57000 km~s$^{-1}$). This event is synchronized with the sharp decay, within less than half a day, of a prominent soft X-ray/UV flare. Accounting for the look-elsewhere effect, the XRISM Resolve data alone indicate a low probability of $2 \times 10^{-5}$ that this UFO detection is due to random chance. The UFO features narrow H-like and He-like Fe lines with a velocity dispersion of $\sim 1000$ km~s$^{-1}$, suggesting that it originates from a dense clump. Beyond this primary detection, there are hints of weaker outflow signatures throughout the rise and fall phases of the soft flare. Their velocities increase from 0.05$c$ to 0.3$c$ over approximately three days, and  they may be associated with a larger stream in which the clump is embedded. The radiation pressure is insufficient to drive the acceleration of this rapidly evolving outflow. The observed evolution of the outflow kinematics instead closely resembles that of solar coronal mass ejections, implying magnetic driving and, conceivably, reconnection near the accretion disk as the likely mechanisms behind both the UFO launch and the associated soft flare. 
}

\authorrunning{L. Gu et al.}
\titlerunning{Delving into the depths of NGC 3783 with XRISM. III.}
\maketitle

\nolinenumbers
\section{Introduction}
\label{intro}
Powerful, highly ionized winds with velocities of 0.1-0.3 times the speed of light and substantial column densities (\NH $\sim 10^{23}$ cm$^{-2}$) 
have been detected in the X-ray and UV spectra of luminous active galactic nuclei (AGNs; \citealt{2003MNRAS.345..705P, 2003ApJ...593L..65R, 2010A&A...521A..57T, 2015ARA&A..53..115K,2015Sci...347..860N,2025Natur.641.1132X}). These so-called ultrafast outflows (UFOs) are primarily identified as highly blueshifted absorption lines in the X-ray spectra of quasars, and some studies indicate their presence in Seyfert galaxies as well \citep{2011MNRAS.413.1251P, 2013MNRAS.430.1102T}. With kinetic energies reaching up to 10$^{46}$~erg~s$^{-1}$, UFOs may be the drivers of AGN feedback, a process in which the central supermassive black hole transfers energy to its host galaxy on a large scale. Research suggests that outflows with mechanical energies exceeding 0.5-5\% of the bolometric luminosity have the potential to expel gas and dust from the host galaxy, thereby suppressing star formation and explaining the observed $M-\sigma$ relation \citep{hopkins2010, 2012ARA&A..50..455F, 2015ARA&A..53..115K}. To better understand this feedback mechanism, it is crucial to observationally constrain the momentum and energy of UFOs and link these to larger-scale molecular outflows on kiloparsec scales.

Despite the increasing number of detections of X-ray UFOs, their geometry and formation mechanisms remain poorly understood. Plausible launching mechanisms include thermal driving \citep{1983ApJ...271...70B, 1993ApJ...402..109B, 2005ApJ...625...95C}, radiation driving \citep{2009PASJ...61L...7O, 2015MNRAS.446..663H, 2018MNRAS.476..512I}, magnetic driving \citep{2010ApJ...715..636F, 2017NatAs...1E..62F}, or a combination thereof. In many sub-Eddington sources with UFOs, the opacity is insufficient for radiative acceleration due to the high ionization state of the outflowing material. Magnetohydrodynamic (MHD) winds there offer a natural explanation for the high velocities of highly ionized material that does not involve line driving. An ultimate, comprehensive understanding of UFO launches and acceleration necessitates a systematic study of their properties (ionization, velocity, and energetics) in relation to fundamental AGN properties like luminosity, accretion rate, and black hole mass \citep{2012MNRAS.422L...1T}. High-resolution X-ray spectroscopy surveys of a broad sample of UFOs are essential for such an analysis.

Before such a comprehensive sample can be assembled, it is essential to determine other properties of UFOs. The equivalent widths and velocities of UFOs  vary significantly on timescales down to hours. These variations can arise from the response of the wind to luminosity changes, or be intrinsic to the launching mechanism itself. The most useful observation would be of a newly formed UFO following an intrinsic change in the accretion structure. Utilizing \textit{XMM-Newton} European Photon Imaging Camera (EPIC) data, \citet{2019MNRAS.484.4287G} presented an intriguing case in Markarian 335, where a potential UFO with a velocity of 0.12$c$ appeared after a flare event triggered by a structural change in the corona. The flare increased the radiation pressure by a factor of 5, which is potentially sufficient to launch the outflow. However, the limited spectral resolution with the CCD instruments and the transient nature of such events have generally hampered high-confidence UFO detections in this and the many other attempts made so far.

High-resolution X-ray spectroscopy provided by Resolve \citep{ishisaki2025, kelley2025resolve} on board XRISM \citep{2025PASJ..tmp...28T} represents a significant advance for studying UFOs. The micro-calorimeter can fully resolve Fe K-shell transitions in the energy band of interest, revealing the velocity and ionization structures of the outflows. Furthermore, time-resolved spectroscopy will allow the characterization of how UFOs respond to variations in the primary emission components of the AGN. Recent observations of the persistent UFO in PDS 456 revealed an absorption line profile with multiple narrow components, indicating clumpiness within a spherically outflowing wind \citep{2025Natur.641.1132X}. In addition, a potential detection of time-varying UFOs in NGC 4151 has been reported by \citet{cindy2025} using the Resolve data.

In this work we explore the presence of UFOs in NGC~3783 by utilizing a recent observation campaign involving XRISM. NGC~3783 is a Seyfert 1 galaxy that hosts an AGN powered by a black hole of mass $2.8 \times 10^{7}$ $M_{\odot}$ \citep{bentz2021} and an Eddington accretion rate of 0.07 \citep{2007MNRAS.378..649S}. This AGN is well known for its prominent, semi-stable warm absorber outflows, characterized by narrow absorption lines with column densities of up to $10^{22}$ cm$^{-2}$ and outflow velocities typically in the range $500-1500$ km s$^{-1}$ \citep{2001ApJ...554..216K, 2002ApJ...574..643K, behar2003, mehd2017, 2019A&A...621A..99M, gu2023}. Obscuring outflows with higher column densities (up to $10^{23}$ cm$^{-2}$) and faster velocities (around $2000$ km s$^{-1}$) have also occasionally been detected. The search for even faster outflows has led to mixed conclusions \citep{2020MNRAS.493.1088I}, as X-ray instruments prior to Resolve lacked the energy resolution and sensitivity to reliably detect absorption lines above 7~keV, especially those that vary on timescales shorter than a day.

This paper is arranged as follows. In Sect.~\ref{sec2} we describe the 2024 observational campaign and the search for UFOs. The properties of the possible outflows are discussed in Sect.~\ref{sec3} and summarized in Sect.~\ref{sec4}. Throughout the paper, the errors are given at a 68\% confidence level. 

\section{Data analysis and results}
\label{sec2}

\subsection{2024 campaign}

A 10-day observing campaign targeting NGC~3783 was carried out in late July 2024, led by XRISM, the new JAXA/NASA/ESA X-ray mission launched in September 2023. XRISM carries two primary instruments: the Resolve X-ray microcalorimeter spectrometer, offering an unprecedented energy resolution of 4.5 eV at 6 keV, and Xtend \citep{noda2025}, an X-ray CCD imager that, while having lower energy resolution, provides better photon-collecting efficiency. Additional instruments participating in the coordinated campaign include the Reflection Grating Spectrometer (RGS), EPIC MOS and pn on board {\it XMM-Newton}, the High Energy Transmission Grating Spectrometer (HETGS) on {\it Chandra}, as well as {\it NuSTAR}, {\it Swift}, NICER, and the Cosmic Origins Spectrograph (COS) on board the {\it Hubble} Space Telescope. In this study, we mainly utilized data from XRISM, {\it XMM-Newton}, and {\it NuSTAR}.

Data reduction was performed using standard pipelines and procedures, in accordance with the cross-calibration paper \citep[Paper II]{xrism2025} for this campaign. Details are not included here. The above paper also provides effective area correction factors for cross-calibration between XRISM, {\it XMM-Newton}, and {\it NuSTAR}, which have been applied in the current analysis.

Apart from Paper II of the series addressing cross-calibration issues from the 2024 observational campaign, the emission and absorption features of the time-averaged Resolve spectrum were reported in Paper I \citep{mis2025}. That work provided a detailed characterization of the quasi-steady, highly ionized outflows in NGC~3783. A total of six outflow components were resolved, including five typical warm absorbers with velocities ranging from 500 to 1200~km~s$^{-1}$, while the sixth component features a broad absorption dip with an outflowing velocity of 14300~km~s$^{-1}$ (0.05$c$). Building on these existing results, this paper performs a search for time-variable, fast outflows using time-resolved spectra.

\subsection{Soft X-ray flare}

\begin{figure*}
    \centering
    \includegraphics[width=0.95\linewidth]{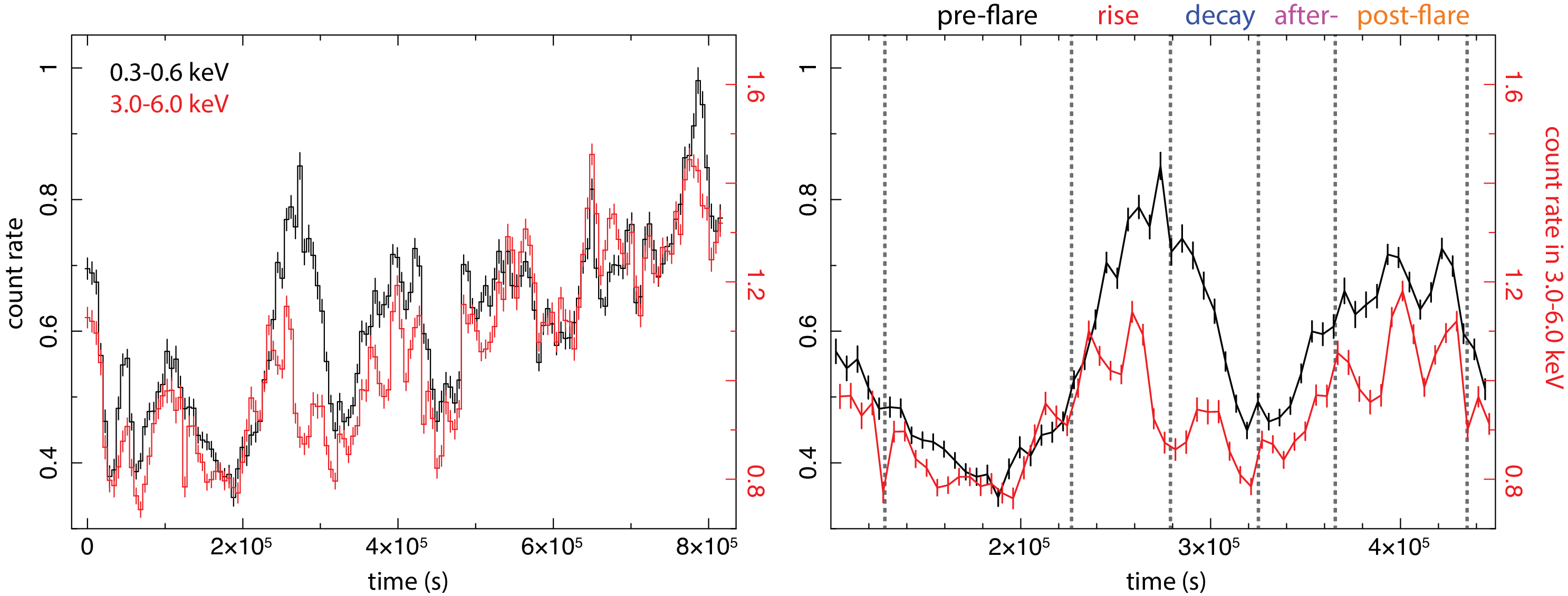}
    \caption{XRISM Xtend light curves from the NGC~3783 campaign. Left: Soft- and hard-band light curves, shown in black and red, respectively. The light curve has been binned to multiples of the XRISM orbit (5747~s), and in this paper we count time since the start of the XRISM observation. Right:  X-ray variability surrounding the main soft flare at $t \sim 2.8 \times 10^5$~s. }
    \label{fig:lc}
\end{figure*}

\begin{figure}
    \centering
    \includegraphics[width=0.98\linewidth]{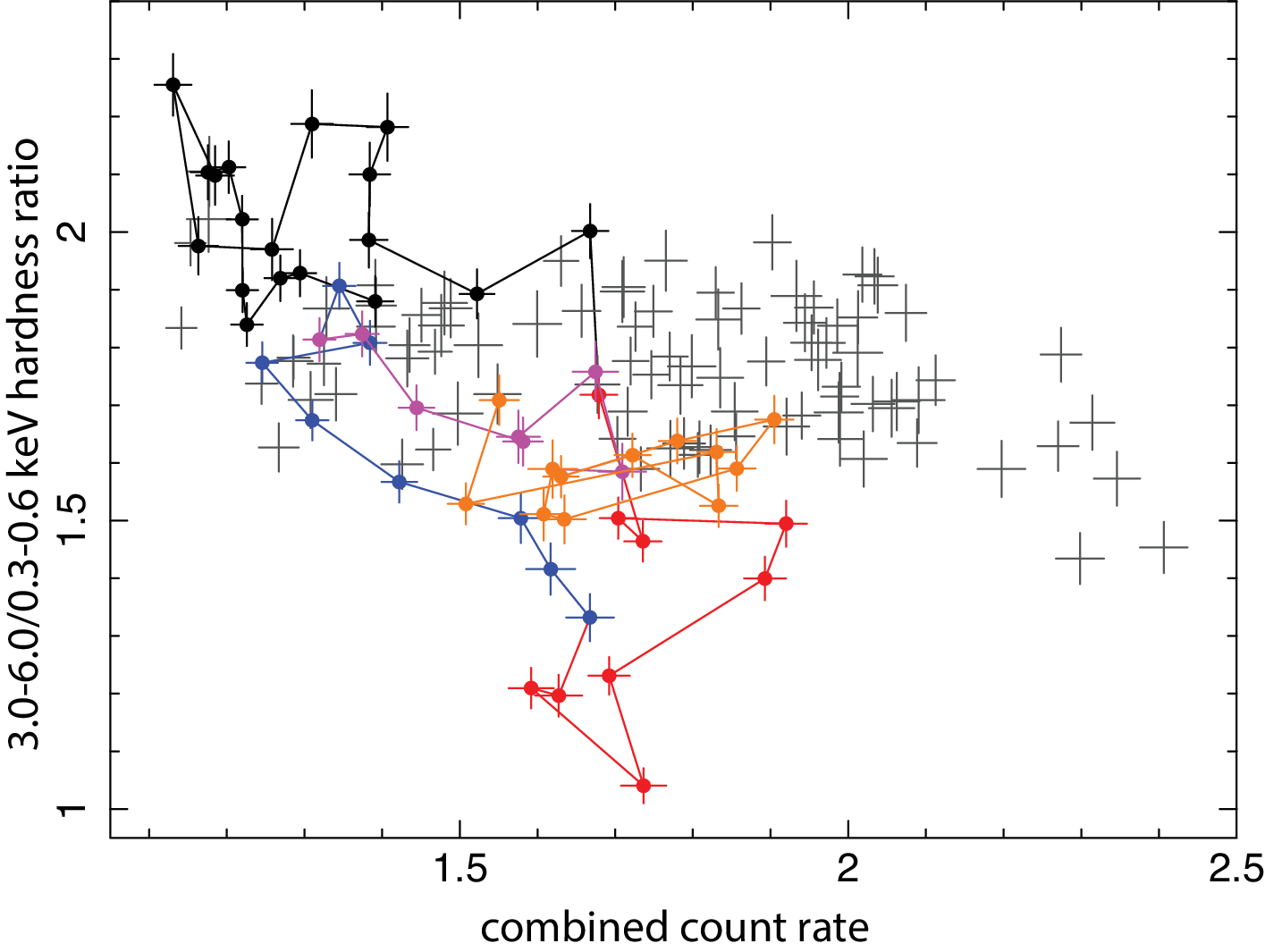}
    \caption{Hardness ratio, defined by the count rates in the $3.0-6.0$~keV and $0.3-0.6$~keV bands, plotted against their combined  count rate. Each data point represents a single XRISM orbit (5747~s). The data are color-coded by flare phase: pre-flare (black), rise (red), decay (blue), after-flare (magenta), and post-flare (orange). Gray points indicate observations outside the main soft flare.}
    \label{fig:hr}
\end{figure}

The Xtend light curves in the $0.3-0.6$ keV and $3.0-6.0$ keV energy bands were extracted to track changes in the soft and hard X-ray continua. These energy bands are selected to avoid known major emission and absorption features, such as the Fe unresolved transition array, Fe L-shell transitions, and the neutral Fe~K$\alpha$ line. The X-ray light curves shown in Fig.~\ref{fig:lc} display a series of significant variability patterns during this campaign. Over the ten days, the X-ray flux increases by about 60\% in both the soft and hard X-ray bands, with multiple flares recurring on shorter timescales of around 150~ks ($\sim 1.7$ days). Among these variations, the period between $1.5 \times 10^{5}$~s and $4 \times 10^{5}$~s clearly needs particular attention. The hard X-ray intensity rises at $t = 2 \times 10^{5}$~s, synchronized with a nearly simultaneous increase in the soft X-ray band. While the hard X-rays peak at $2.6 \times 10^{5}$~s and begin to decline, the soft X-rays continue to increase, peaking approximately $4 \times 10^{4}$~s later. With the signature feature being a peak in soft X-rays, we refer to it as the ``soft flare'' hereafter. A secondary soft flare-like event appears toward the end of the campaign; however, since the second peak is only partially covered by the campaign, our analysis will focus on the primary soft flare.

To investigate the spectral variation in the soft flare, we divided the event into five periods as illustrated in Fig.~\ref{fig:lc} and Table~\ref{tab:Felines}. The net Resolve exposure time is approximately 25~ks for the rise, decay, and after-flare phases, 35~ks in the post-flare phase, and about 50~ks in the pre-flare phase. 

Figure~\ref{fig:hr} displays the evolution of the hardness ratio during the soft flare event. The pre-flare phase begins with the hardest spectrum observed throughout the campaign, corresponding to the lowest overall luminosity. During the rise phase, the source transitions to a high state while exhibiting the softest spectrum recorded during the campaign. Throughout the decay and post-flare phases, the source bounced between harder and softer states before gradually relaxing to a stable hardness ratio in the post-flare phase. This flare event clearly exhibits the largest dynamical range in the hardness ratio, a striking contrast to the relatively stable level observed in the rest of the campaign.

\subsection{Outflow associated with the soft flare event}

\begin{figure}
    \centering
      \includegraphics[width=0.9\linewidth]{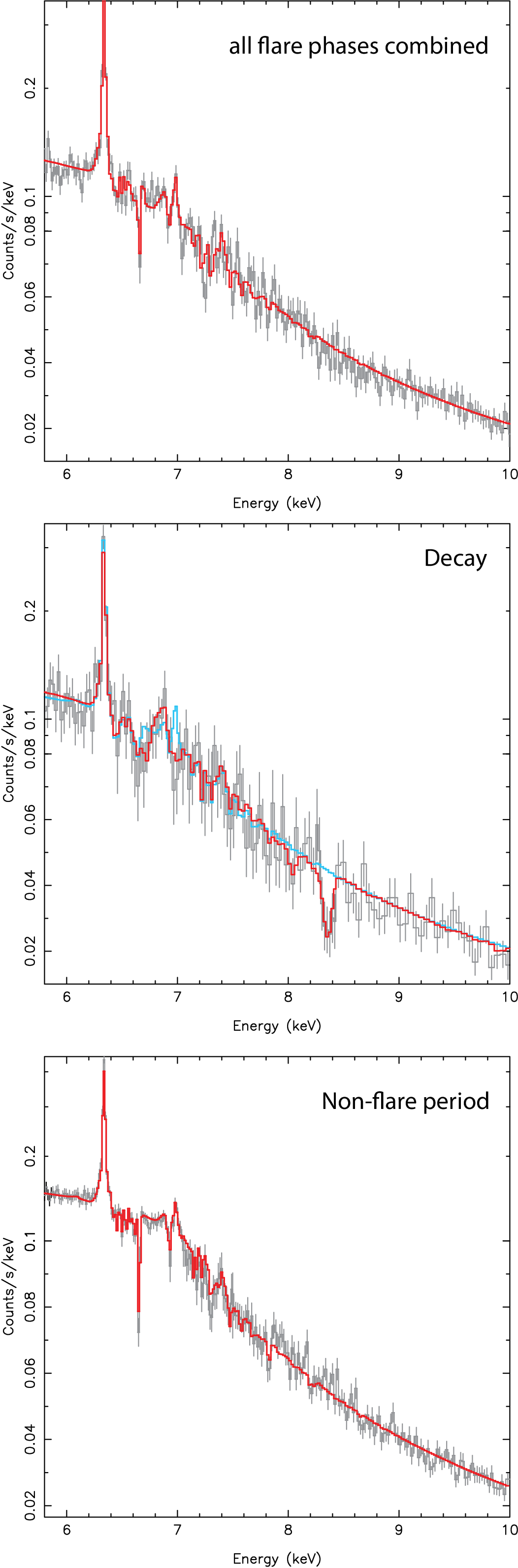}
    \caption{Time evolution of the NGC~3783 Resolve spectra. Top: Average spectrum over all flare phases. Middle: Spectrum during the decay phase only. Bottom: Average spectrum combining all periods outside the soft flare, shown for comparison. The dip at $\sim 8.4$ keV is unique to the decay phase. The light blue curve in the middle panel shows the best-fit model excluding both the UFO and the 3700~km~s$^{-1}$ components. }
    \label{fig:decayonly}
\end{figure}

The Resolve spectrum during the soft flare period, from $t = 1.3 \times 10^{5}$~s to $4.3 \times 10^{5}$~s, is analyzed based on the phases as defined in Fig.~\ref{fig:lc} and Table~\ref{tab:Felines}.

\subsubsection{General spectral components applied to all phases}

To search for spectral features associated with the soft flare, we modeled the Resolve spectrum in the $4.0-10.0$~keV band based on the following components. The observed continuum was modeled in the same way as Paper I. A power-law component fits the primary continuum, with the photon index $\Gamma$ obtained to be $1.79 \pm 0.01$. To account for potential soft X-ray excess, we included a warm Comptonization component, {\tt comt}. Although the contribution of {\tt comt} to the Resolve band is minor, $\approx 2$\%, it remains essential as a spectral energy distribution (SED) component for photoionization calculations. Since the {\tt comt} parameters cannot be constrained with Resolve data, we fixed them to those of the 2001 unobscured model from \citet{mehd2017}. The power-law and {\tt comt} components together form the SED used for photoionization calculations. In our time-resolved analysis, we allowed the normalization of both components and the photon index of the power-law component to vary freely.

Our model includes relevant narrow emission line components: \FeKa, \FeKb, Ni~K$\alpha$, \lya, and \lyb. These components are incorporated using Gaussian lines, following Paper I, including the line widths. We also considered the broad emission component, including the nonrelativistic component from the broad line region, as well as a relativistic component from the inner accretion disk. The broad line component is handled with a broad Gaussian and the relativistic component is accounted for with the \FeKa convolved with the {\tt spei} relativistic line profile component. These components are added in the same way as Paper I and Li et al. in prep.

The low-velocity ionized absorbers are modeled with the {\tt pion} components \citep{mehd2016, miller2015}, in the same way as Paper I and Zhao et al. in prep. These components are found to be quasi-stable \citep{gu2023}, and our analysis is not sensitive to minor variability within these warm absorbers. The {\tt pion} components are fed with the year 2001 unobscured SED from \citet{mehd2017}, which is consistent with the intrinsic UV and X-ray continuum of our observation. The column density, ionization parameters, and kinetic properties of these components are fixed to the values reported in Paper I, which agrees with the full absorption measurement analysis in Zhao et al. in prep. The very high-velocity component X from Paper I, with an outflow velocity of about 0.05$c$, is not included as a general component in the baseline model, as it already qualifies as a UFO candidate. As shown later in Sect.~\ref{sec:otherufo}, the component X likely exhibits significant variability.

As shown in Fig.~\ref{fig:decayonly}, this model fits the time-average spectrum well during the entire soft flare period. It is established as our baseline, defining the starting point for the subsequent analysis.

\begin{figure*}
       \centering
    \includegraphics[width=0.98\linewidth]{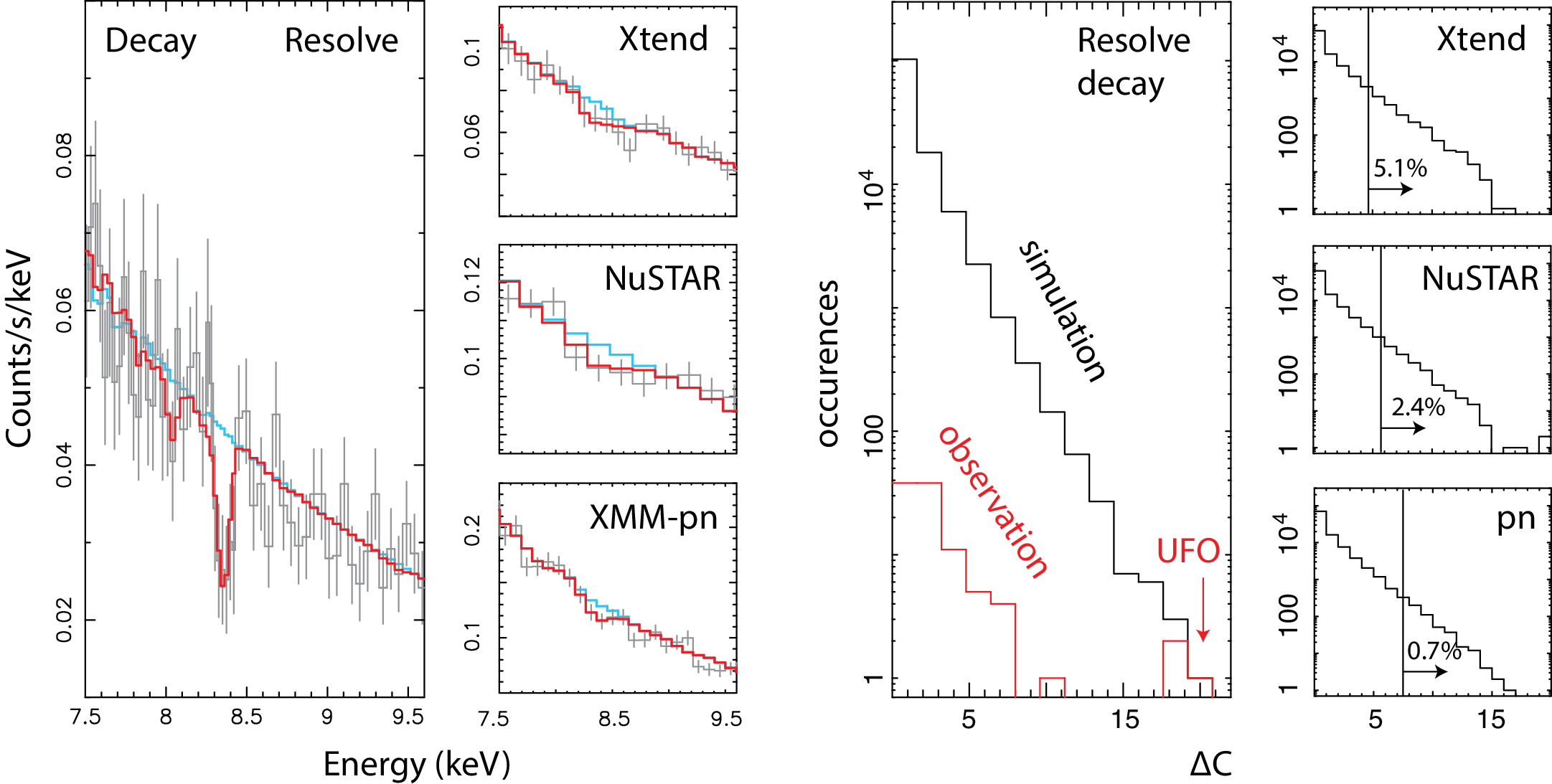}
    \caption{Detailed look at the dip at 8.4~keV and the LEE. Left: Resolve, Xtend, \textit{NuSTAR}, and \textit{XMM-Newton} pn spectra from the decay phase. The red line shows the best-fit model to the Resolve data and including the UFO component, while the light blue line represents the model without it. Right: $\Delta C$ distribution from $1.3 \times 10^{5}$ runs of the Monte Carlo simulation of the Resolve spectrum, and $1 \times 10^4$ times for each of other three instruments (black histograms) and the observed $\Delta C$ distribution for the Resolve data (red histograms). Vertical lines mark the observed $\Delta C$ values for the other three instruments.}
    \label{fig:spec2}
\end{figure*}

\subsubsection{Possible detection of an ultrafast outflow}
\label{decayufo}

The most striking feature, as shown in Fig.~\ref{fig:decayonly}, is an absorption dip centered at 8.4~keV during the decay phase. To characterize this feature, we incorporated a {\tt pion} component on the baseline model. The power-law and {\tt comt} components serve as the input SED, while the ionization parameter, column density, bulk velocity, and velocity dispersion are allowed to vary for the {\tt pion} component. The warm absorbers and emission line components in the baseline model remain fixed at their time-averaged best-fit values, since their potential variability did not affect the observed dip. The results from the spectral analysis are summarized in Table~\ref{tab:Felines}.

 Modeling the dip at 8.4~keV using a Gaussian absorption line reveals a C-stat improvement of 17. Employing the $\tt pion$ model yields a slightly enhanced $\Delta C$ of 20. This further improvement is probably caused by the fact that $\tt pion$  tends to include both H- and He-like Fe lines, accounting for the 8.4~keV dip as well as a secondary absorption feature near 8~keV. The best-fit ionization parameter (log$\xi$), column density, outflow velocity, and velocity dispersion are 3.2$\pm 0.2$, 1.1$\pm 0.4 \times 10^{23}$ cm$^{-2}$, 56780$\pm 450$ km s$^{-1}$, and 1170 $\pm 350$ km s$^{-1}$. The outflow velocity indicates a UFO speed of $0.19c$ ($\sim 57000$ km~s$^{-1}$).

The instrumental background of Resolve cannot explain the 8.4~keV dip. No instrumental features overlap with this energy, and the energies of the nearest background lines, including Cu~K$\alpha$ at 8.03 and 8.05 keV and Au~L$\alpha$ at 9.71 and 9.63 keV, are too different to explain the feature.

To improve confidence in our results, we cross-checked the detection with data from other instruments. The decay phase is well covered by Xtend, \textit{XMM-Newton}, and \textit{NuSTAR}. For these instruments, we conducted data reduction following the procedure described in the cross-calibration paper \citep{xrism2025}. As shown in Fig.~\ref{fig:spec2}, the spectra from Xtend, \textit{NuSTAR}, and \textit{XMM-Newton} pn were modeled using the same baseline model as Resolve, while allowing the normalization and photon index of the power-law component to vary as to address cross-calibration uncertainties \citep{xrism2025}. In the same plot, we also show the model incorporating the best-fit {\tt pion} component from Resolve. The parameters of the {\tt pion} component are fixed to the Resolve values. While the lower spectral resolution of Xtend, \textit{NuSTAR}, and \textit{XMM-Newton} pn prevents the narrow feature from being fully resolved, the comparisons demonstrate a statistical preference with the Resolve solution. The inclusion of the pion component results in improvement of C-stat of 4.7, 6.8, and 7.6, with respect to the baseline model, for Xtend, \textit{NuSTAR}, and \textit{XMM-Newton} pn. The combined improvement in C-stat across all four instruments, including Resolve, is 39.1.

The $\Delta C$ values reported above cannot be directly used to determine the confidence level of the detected feature. This limitation arises because, given the energy range explored during the analysis, the look-elsewhere effect (LEE) associated with detecting a random feature cannot be neglected. To address the LEE, we employed a Monte Carlo simulation approach. Following the methodologies outlined by \citet{2018MNRAS.473.5680K} and \citet{2020MNRAS.492.4646P}, we first simulated a large set of spectra for each instrument, including 1.3 $\times 10^{5}$ times for Resolve, and 1 $\times 10^{4}$ for the others. The numbers of simulations correspond to the p-values associated with $\Delta C$ = 20 and 10, respectively.

The simulations were done with the baseline model for the decay phase and utilizing the actual exposure time. For each simulated spectrum, we scanned the $7.5-9.0$ keV range with the {\tt pion} model for the UFO and computed the $\Delta C$ relative to the baseline. Since no emission line was detected in the actual data, the scan did not include searches for emission features. The distribution of $\Delta C$ values from these simulations, representing the likelihood of random features, is shown in Fig.~\ref{fig:spec2}.

For Resolve, the occurrence $N$ can be described as a function of $\Delta C$ by $N = 10^{-0.23 \Delta C + 4.67}$. Integrating this function over the range of observed $\Delta C$ = 20 to infinity results in an expected frequency of $\sim 2 \times 10^{-5}$. This frequency corresponds to an effective $\Delta C$ of 18. This suggests that the LEE would reduce the $\Delta C$ by 2 for Resolve. As shown in Fig.~\ref{fig:spec2}, the same exercise has been done for other instruments, and the effective $\Delta C$ of 3.8, 5.1,and 7.1 for Xtend, \textit{NuSTAR}, and pn data, reducing 0.9, 1.7, and 0.5 from their original values. The combined effective $\Delta C$ is 34, leading to a random probability of $3 \times 10^{-7}$ after accounting for the LEE.

\subsubsection{Potential outflow at 3700 km s$^{-1}$}

Modeling the Resolve spectrum in the decay phase with the baseline model reveals another feature: the narrow Fe K$\beta$ fluorescence line at approximately 7~keV appears to be absent. As shown in Fig.~\ref{fig:spec}, this line is present during other phases of the flare, indicating that the narrow band at 7 keV varies on a timescale of less than or equal to 40 ks. To investigate whether this variability is due to intrinsic changes in the emission component, we allowed the relevant emission components to vary, including the narrow Fe~K$\beta$ line and the broad relativistic Fe line, which likely exhibits a cutoff around 7~keV (Li et al., in prep.). For this relativistic component, both the normalization and emissivity slope are set free. The new fit constrained the Fe K$\beta$/Fe K$\alpha$ ratio to an upper limit of 0.035, significantly lower than the value with the baseline model of 0.12. This drastic change cannot be caused by the narrow line variation because there is no physical solution for such a line ratio change with a stable line center \citep{yamaguchi2014}, and the observed 40~ks timescale variability is incompatible with a distant reflector origin of the narrow line.

The remaining possibility is that an absorption feature from an outflow overlaps in energy with the narrow Fe~K$\beta$ line. To address this, we incorporated two {\tt pion} components into the baseline model; one for the known outflow at $0.19c$ and another for the absorption at Fe~K$\beta$. The best-fit parameters for the new {\tt pion} component are as follows: log$\xi$ = $2.9 \pm 0.1$, column density = $5.9 \pm 1.3$ $\times 10^{22}$  $\text{cm}^{-2}$, outflow velocity = $3720 \pm 480$ $\text{km s}^{-1}$, and velocity dispersion = $1470 \pm 530$ $\text{km s}^{-1}$. In fact, this additional component absorbs the continuum around 7~keV, causing the Fe~K$\beta$ line to appear as if it is missing, because the emission line happens to ``fill in'' the absorption dip in the continuum.

Including this component improves the C-statistic by 19. Taking into account the LEE, in the same way as for the UFO component (Sect.~\ref{decayufo}), yields a random probability of approximately $3 \times 10^{-5}$. The outflow velocity associated with this component is significantly higher than that of the normal warm absorbers in NGC~3783, yet lower than the UFO components. Looking at past observations, another transient outflow component that also obscured the Fe~K$\beta$ line is the high-ionization component reported in \citet{mehd2017} using {\it XMM-Newton} CCD data. This component had a velocity of about 2300~km s$^{-1}$ and a column density of $1-2 \times 10^{23}$ $\text{cm}^{-2}$, identified during an obscuration event in 2016.

The 3700~km~s$^{-1}$ component appears to be transient, possibly related to the launch of the UFO at 0.19$c$ ($\sim 57000$ km~s$^{-1}$). As this paper focuses on the UFO component, the nature of 3700~km~s$^{-1}$ component will be discussed in a separate paper.

\subsubsection{Search for outflows in other phases}
\label{sec:otherufo}

\begin{figure}
    \centering
    \includegraphics[width=0.98\linewidth]{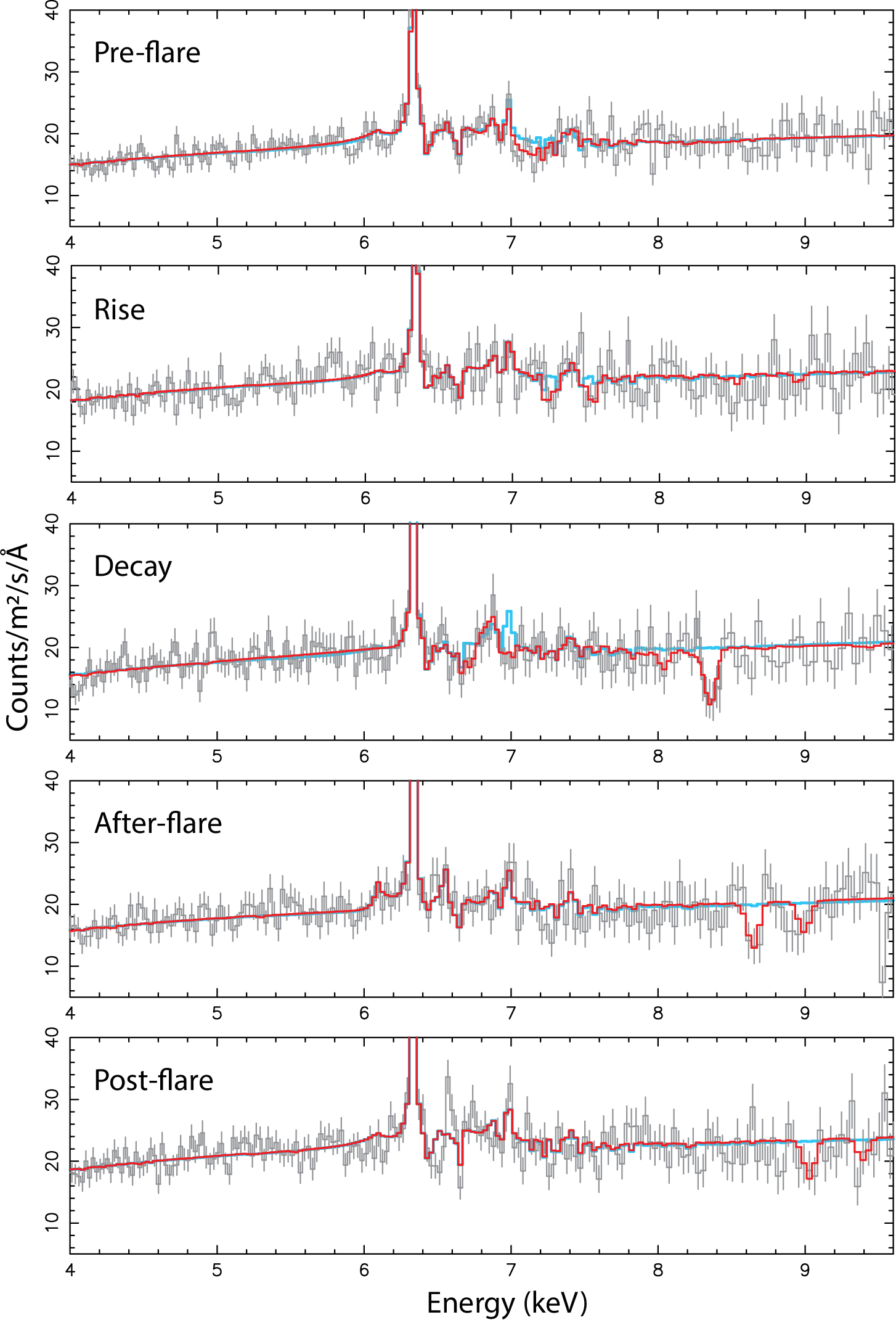}
    \caption{Phase-resolved Resolve spectra during the soft flare. The best-fit models are shown in red. The light blue curves show the model that excludes both the UFO and the 3700~km~s$^{-1}$ component.}
    \label{fig:spec}
\end{figure}

\begin{table*}[htpt!]
 \caption{Results from fitting the outflows.}
 \centering
  \begin{tabular}{lccccc||c}
   \hline
   Parameters & Pre-flare & Rise & Decay$_{\rm 1}$$^{a}$ & After & Post-flare & Decay$_{\rm 2}$$^{a}$ \\

   \hline
    $t_{\rm start}$$^{b}$ ($10^{5}$ s)   & 1.3 & 2.3 & 2.8 & 3.2 & 3.6 & 2.8 \\ 
    $t_{\rm end}$$^{b}$ ($10^{5}$ s)     & 2.3 & 2.8 & 3.2 & 3.6 & 4.3 & 3.2 \\ 
   \logxi (erg~cm~s$^{-1}$) & 2.7$\pm 0.1$  & 3.3$\pm 0.1$  & 3.1$\pm 0.2$  & 2.8$\pm 0.1$ & 2.7$\pm 0.1$ & 2.9$\pm 0.1$ \\
   \NH ($10^{22}$~cm$^{-2}$) & 3.8$\pm 2.2$ & 1.8$\pm 0.8$ & 10.9$\pm 2.7$ & 3.2$\pm 1.3$ & 1.8$\pm 0.9$ & 5.9$\pm 1.3$ \\
   \vout (km s$^{-1}$) & 14310$\pm 3870$ & 26600$\pm 640$ & 56780$\pm 450$ & 77600$\pm 560$ & 89620$\pm 650$ & 3720$\pm 480$ \\
   \sigv (km s$^{-1}$) & 4570$\pm 1950$ & 730$\pm 460$ & 1170$\pm 350$ & 1240$\pm 490$ & 1020$\pm 640$ & 1470$\pm 530$ \\
   \hline
   C-stat  & 655 & 690 & 689 & 534 & 609 & 670 \\
   Expected C-stat & 624$\pm 32$ & 634$\pm 33$ & 643$\pm 33$ & 481$\pm 31$ & 596$\pm 32$ & 643$\pm 33$ \\
   $\Delta C$$^{c}$ & 12 & 10 & 20 & 12 & 9 & 19 \\
   $P_{\rm LEE, Resolve}$$^{d}$ & $7 \times 10^{-4}$ & $2 \times 10^{-3}$ & $2 \times 10^{-5}$ & $7 \times 10^{-4}$ & $4 \times 10^{-3}$ & $3 \times 10^{-5}$ \\
   $P_{\rm LEE, all}$$^{e}$ & $4 \times 10^{-4}$ & $2 \times 10^{-3}$ & $3 \times 10^{-7}$ &  $2 \times 10^{-4}$ & $4 \times 10^{-3}$ &  \\
   \hline
  \end{tabular}
  \label{tab:Felines}
  
      \vspace{0.5em} 
    \tablefoot{$^{a}$ The UFO and the 3700~km~s$^{-1}$ components are distinguished by footnotes 1 and 2.
    $^{b}$ Start and end times of the respective phases.
        $^{c}$ $\Delta C$ value obtained directly from fitting the Resolve spectra.
        $^{d}$ Random probabilities estimated by considering the LEE based on best-fit and simulations using the Resolve data only.
        $^{e}$ Random probabilities estimated as in $d$, but combining best-fits and simulations from all instruments. For the 3700~km~s$^{-1}$ component we use the Resolve data alone.}

\end{table*}

\begin{figure}
    \centering
    \includegraphics[width=0.98\linewidth]{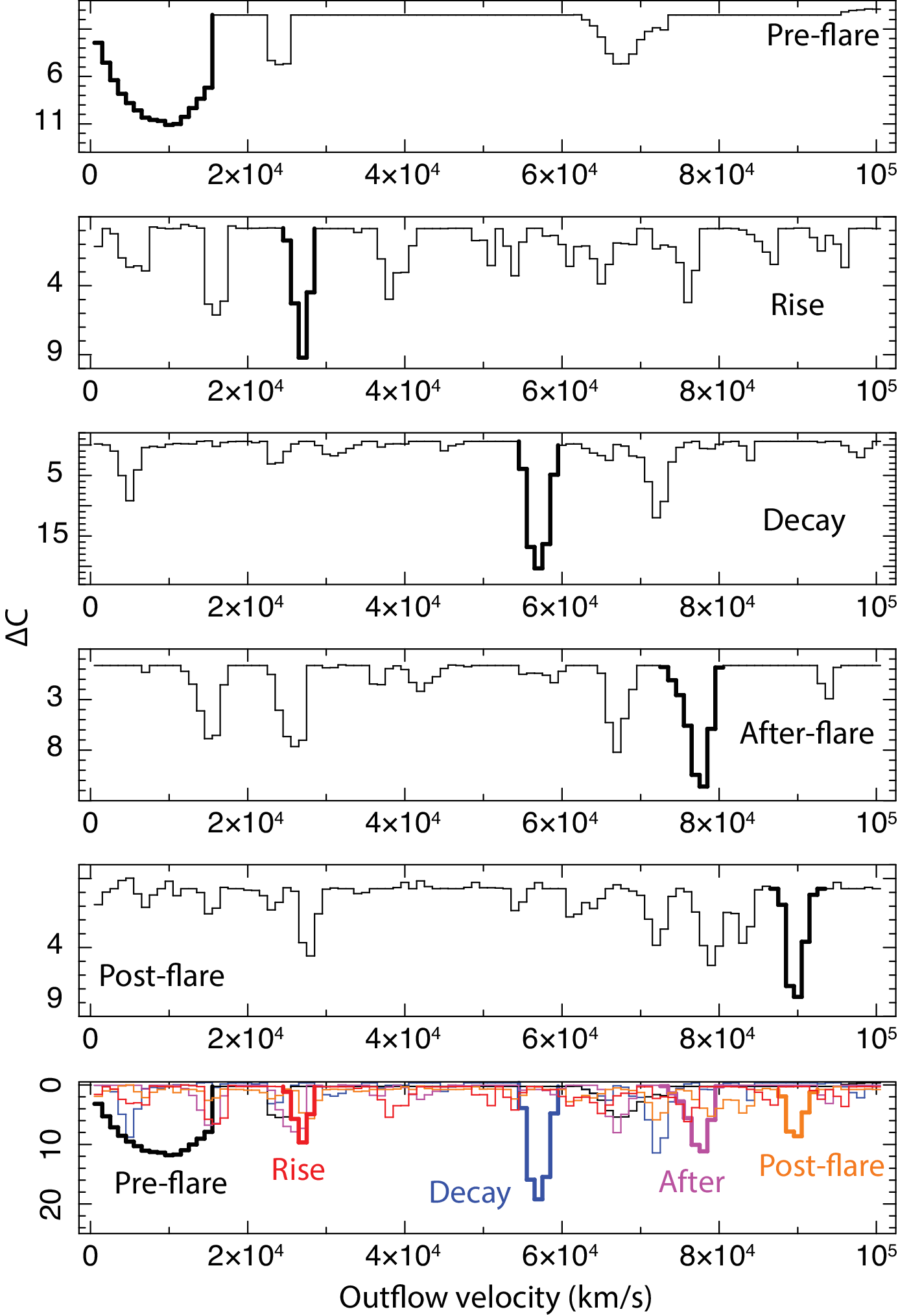}
    \caption{Potential features identified with the {\tt pion} scan on the Resolve data. The curves show the $\Delta C$ obtained in each step of the scan. The most prominent feature in each phase is highlighted with a thick line. In the bottom panel, all individual curves from the upper panels are combined and presented on a common $\Delta C$ scale for comparison. }
    \label{fig:lee}
\end{figure}

\begin{figure}
    \centering
    \includegraphics[width=0.98\linewidth]{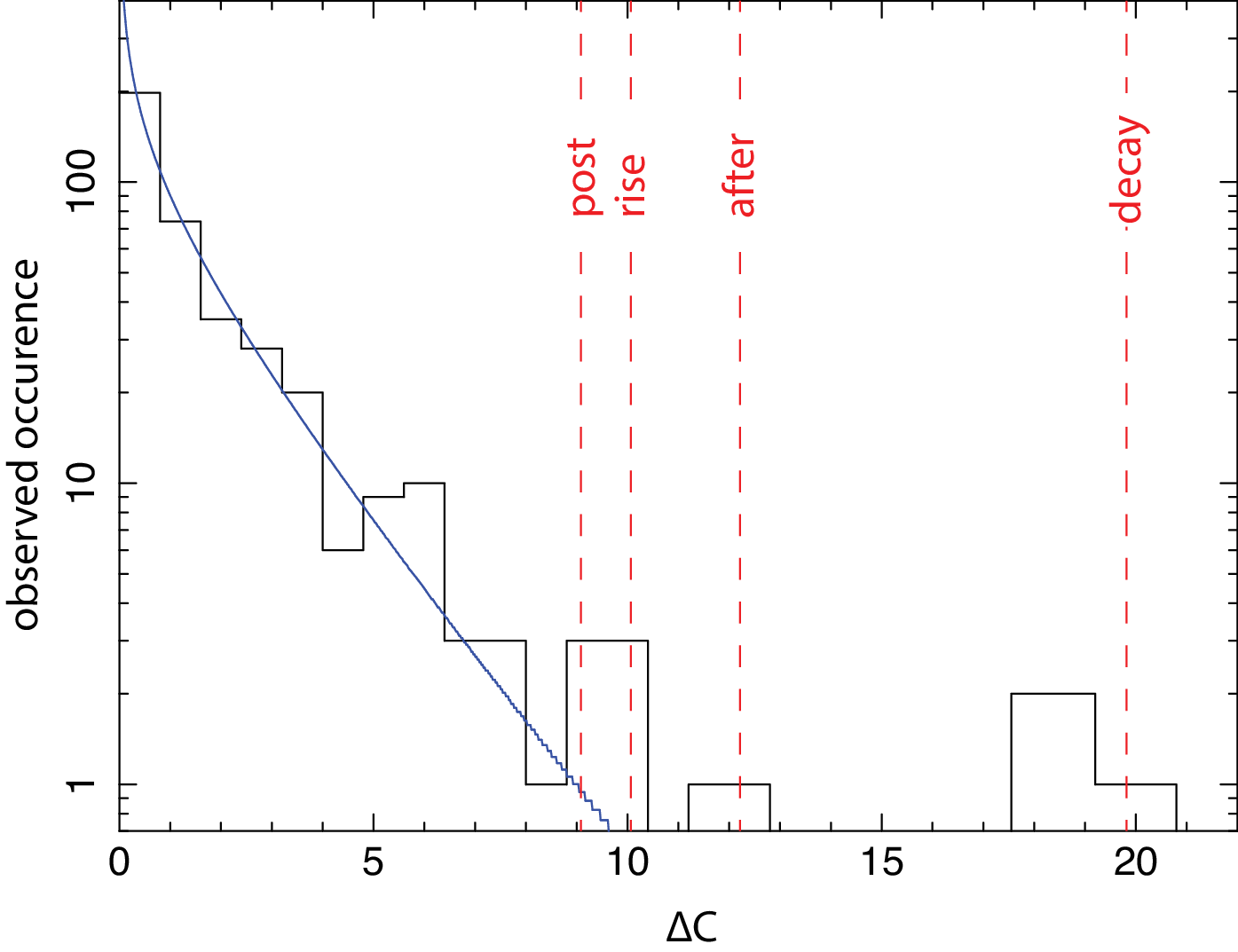}
    \caption{LEE in the {\tt pion} scan. The black histogram shows the combined $\Delta C$ distribution from the rise, decay, after, and post-flare phases. The blue line represents an exponential fit to this histogram, following a form similar to that described in Sect.~\ref{decayufo}. The dashed red lines mark the $\Delta C$ values of the most prominent features identified in these phases. }
    \label{fig:threecomb}
\end{figure}

A visual examination of Fig.~\ref{fig:spec} suggests the potential presence of further absorption features in the other phases of the flare. To systematically search for absorbers, we scanned the phase-resolved spectrum using a {\tt pion} component added to the baseline model. In each iteration ($n$), the bulk velocity of the {\tt pion} component was set to $n \times 1000$ km s$^{-1}$, while its column density, ionization parameter, and velocity dispersion were set as free parameters. The normalization and photon index of the power law, along with the normalization of the {\tt comt} component, were also left free. Meanwhile, the warm absorbers and emission components remained fixed at their time-averaged values.

As shown in Fig.~\ref{fig:lee}, the statistical significance of the {\tt pion} component is plotted against the scanning velocity of $n \times 1000$ km s$^{-1}$. To evaluate the contribution of random variations, we present a comparison in Fig.~\ref{fig:threecomb} between the observed $\Delta C$ diagram and the expected distribution derived from Poisson data. To maximize statistical significance, we combined the rise, decay, after-, and post-flare phases in the comparison. The pre-flare phase was excluded due to its potential broad absorption features, which would have resulted in bins that are not fully independent. The comparison indicates that features detected with $\Delta C$ $< 9$ are likely due to Poisson noise. Peaks with $\Delta$C $\geq 9$ may have a higher likelihood of being real. Their actual significance will be evaluated considering the LEE, which will be detailed in Appendix~\ref{app:lee}. We also attempted to enhance the detection of the potential outflow by combining the Resolve data with those from other instruments. The likelihood presented in Table~\ref{tab:Felines} takes into account the best-fit and the LEE simulations from all instruments. It can be inferred that the CCD instruments show qualitative agreement with the presence of the {\tt pion} component during all phases, though the overall improvement on the detection significance remains modest except for the decay phase.

The relatively large uncertainties in the column densities listed in Table~\ref{tab:Felines} suggest a low confidence level for the fast outflow feature, except during the decay phase. However, we caution that the error on the hydrogen column density may not be an ideal indicator of significance, as it can be highly degenerate with the ionization parameter for a single absorption feature.

As detailed in Table~\ref{tab:Felines}, the pre-flare phase is characterized by a significant outflow with a velocity of 0.05$c$ and a velocity dispersion of 4500~km s$^{-1}$. These kinematic parameters show a good agreement with component X, which was identified in the time-averaged Resolve spectrum and reported in Paper I of the NGC~3783 series \citep{mis2025}.

To enable a direct comparison with component X, we fixed the ionization parameter (\logxi) of the pre-flare component at 4, aligning with the value assumed in Paper I. This yielded a best-fit column density of $1.1 \times 10^{23}$~cm$^{-2}$, which agrees with component X. This indicates that, when accounting for the potential degeneracy in the parameter space between the ionization parameter and the column density, the 0.05$c$ component in the pre-flare is indeed consistent with the component X identified in the time-averaged spectrum.

In the post-flare phase, Fig.~\ref{fig:spec} reveals a potential extra narrow emission component at around 6.6~keV. This feature is well described by a narrow Gaussian with $\sigma \sim 200$ km~s$^{-1}$, improving the C-stat by 18. Since it may form a P-Cygni profile with the adjacent Fe~XXV absorption feature from the warm absorber, this emission could be associated with slow outflows.

Although the statistical uncertainties on the outflow velocity are less than 1000~km~s$^{-1}$ from the rise to the post-flare phases, the systematic uncertainties may be larger. Taking the decay spectrum as an example, the current best-fit solution associates the main absorption dip at $\sim 8.4$~keV with Fe~XXVI, with a secondary feature at $\sim 8$~keV corresponding to Fe~XXV. The overall outflow velocity is then approximately 0.19$c$. However, as shown in Fig.~\ref{fig:lee}, the {\tt pion} scan suggests an alternative possibility with an outflow velocity of 0.24$c$, where the main dip at $\sim 8.4$~keV is instead caused by Fe~XXV absorption. Although this alternative model has a $\Delta C$  value worse by 8 compared to the best-fit, it can be considered a source of systematic uncertainty on the outflow velocities.

In summary, we identify a potential outflow event during the strong soft flare in NGC~3783, with the most significant feature observed at $v \approx 0.19c$ during the decay phase. Furthermore, Resolve spectra suggest the potential presence of outflow signatures in other flare phases as well, exhibiting velocities that rise from 0.05$c$ to 0.3$c$ within a total time span of 3$\times 10^{5}$~s.

\section{Discussion}
\label{sec3}

In this section we closely examine the observed physical properties of the potential outflow and explore the various processes that could potentially explain its observed kinematics.

\begin{figure*}[!htb]
    \centering
    \includegraphics[width=0.98\linewidth]{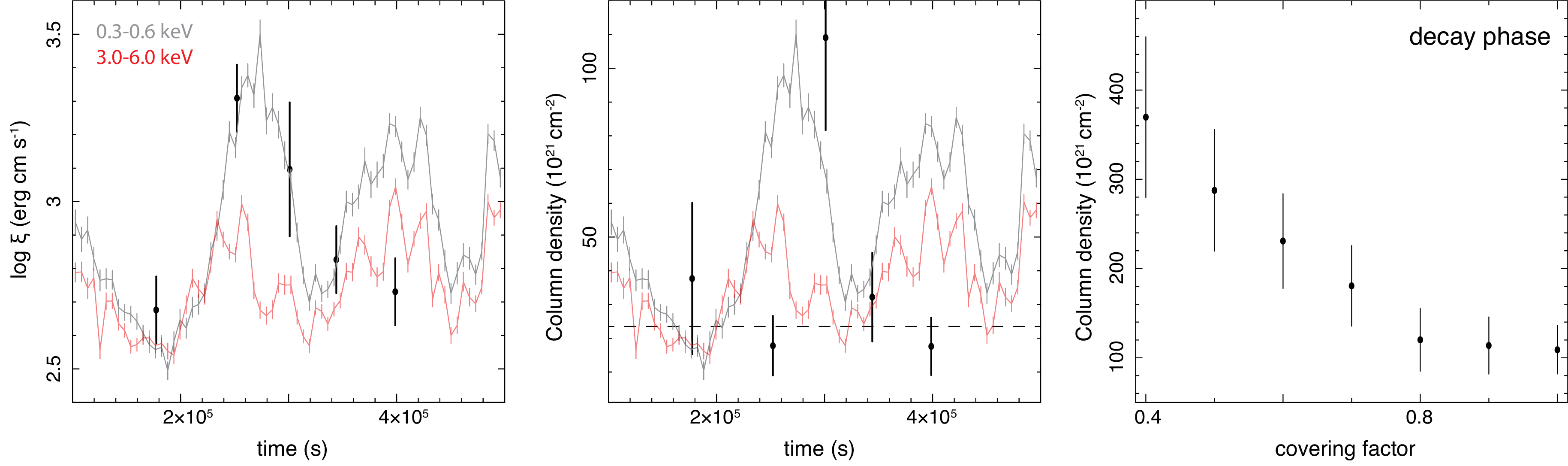}
    \caption{Evolution of outflow parameters during the flare. Left: Outflow parameter. Middle: Column density. These two panels are overlaid with the soft-band (black) and hard-band (red) X-ray light curves. A horizontal dashed line in the middle panel indicates the average column density excluding the decay phase. Right: Column density as a function of the covering factor for the decay phase. }
    \label{fig:xi}
\end{figure*}

\subsection{Constraints on the ionization and column density}

As shown in Fig.~\ref{fig:xi}, the ionization parameter of the UFO exhibits an increase from $2.7 \pm 0.1$ during the pre-flare phase to $3.3 \pm 0.1$ during the rise phase, followed by a smooth decline back to $2.7 \pm 0.1$. Since the continuum luminosity in the $1-1000$ Rydberg range approximately doubles from the pre- to rise phases and then decreases by a factor of 1.7 during the rise to post-flare transition, the observed change in ionization parameter broadly agrees with the flaring luminosity variation. This suggests that the upper limit on the temporal change in the product of density and squared distance ($nR^{2}$) across the flare is approximately a factor of 2.

Figure~\ref{fig:xi} also shows that the observed column density remains relatively stable within the range $2-4 \times 10^{22}$ cm$^{-2}$, except possibly during the decay phase, where it is about $3-5$ times higher than the average. This suggests that while the outflow is generally stable, a denser clump may enter the line of sight during the flare decay. These column densities are consistent with recent XRISM measurements of UFOs in PDS456 \citep{2025Natur.641.1132X}, and higher than the column density of $3 \times 10^{21}$ cm$^{-2}$ found in PG~1211+143 with the {\it Chandra} grating \citep{danehkar2018}. Considering relativistic effects as discussed in \citet{luminari2020}, the actual column densities are likely underestimated; applying their correction implies $\sim 2 \times 10^{23}$ cm$^{-2}$ during decay and around $4 \times 10^{22}$ cm$^{-2}$ otherwise.

A key systematic uncertainty in the column density measurement arises from the covering factor of the outflow component. In our analysis, we assumed full coverage of the X-ray source, but in principle, the outflow could be only partially covering the source. To investigate this, we refit the spectrum by manually varying the covering factor of the {\tt pion} component. As shown in Fig.~\ref{fig:xi}, the column density of the UFO increases from $\sim 1 \times 10^{23}$~cm$^{-2}$ to about $4 \times 10^{23}$~cm$^{-2}$, while the covering factor decreases from 1 to 0.4. The C-stat remains nearly unchanged across this range. A covering factor below 0.4 results in dips that are too shallow to reproduce the observed 8.4 keV feature. This suggests that the projected area of the outflow cannot be much smaller than the X-ray emitting region at this energy.

\subsection{Constraints on the density and location}
\label{dis:density}

\begin{figure*}[!htb]
    \centering
    \includegraphics[width=0.98\linewidth]{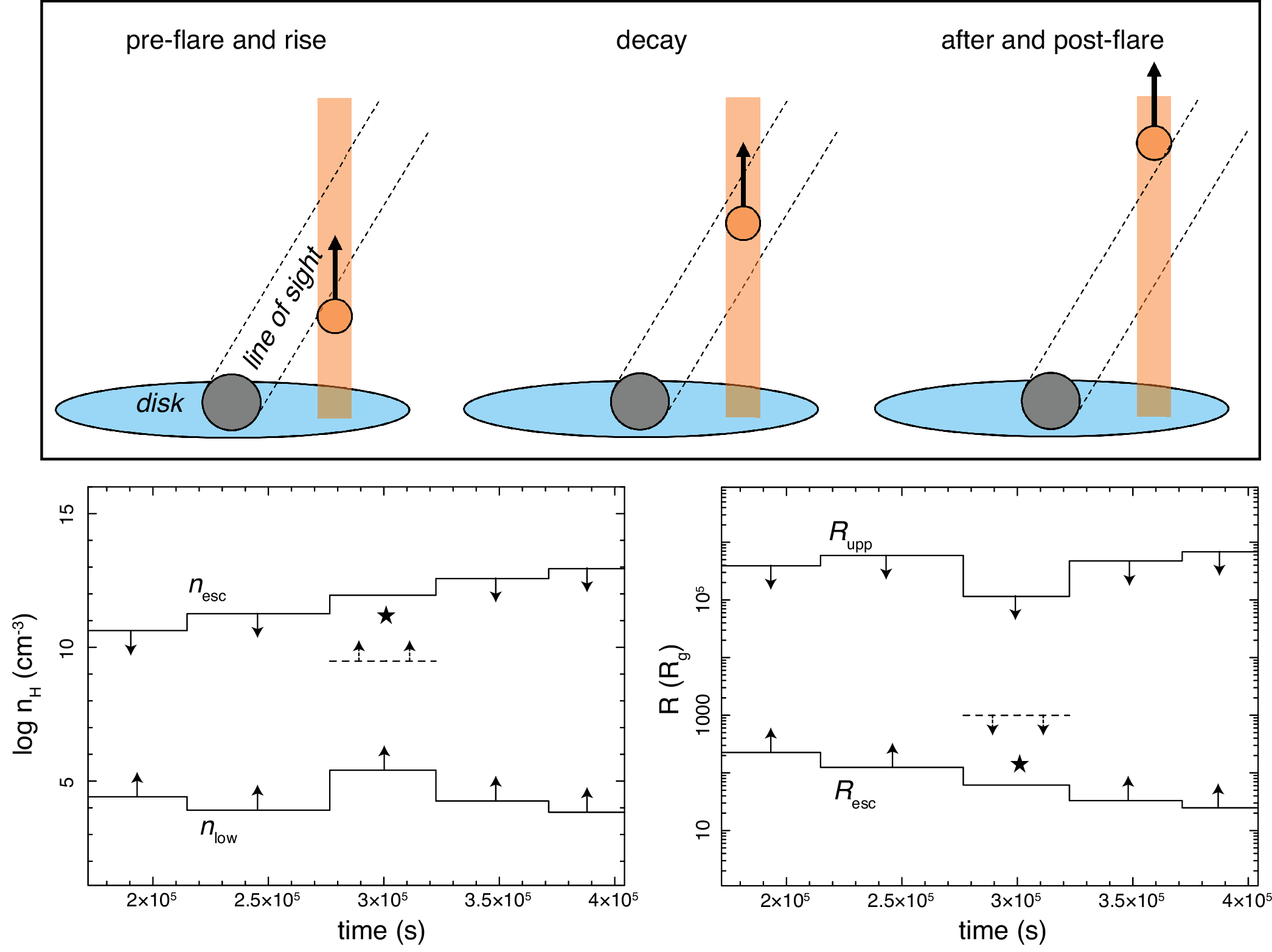}
    \caption{Constraints on the density and distance of the outflow. Top: Schematic illustrating a simple scenario: a clumpy outflow stream moving vertically from near the accretion disk, with the clump represented by the circle. Bottom: Density (left) and distance (right) constraints derived from the geometrical limits ($n_{\rm low}$ and $R_{\rm upp}$) and escape velocity ($n_{\rm esc}$ and $R_{\rm esc}$). Stars mark the density and distance values assuming that a clumpy component, observed during the decay phase, was launched at the pre-flare onset. The dashed line shows the density and distance limits estimated from the duration of the 8.4~keV dip during the decay phase.}
    \label{fig:den_dist}
\end{figure*}

To better understand the physical properties of the outflow, we estimated the density and distance constraints based on geometrical considerations. Assuming that the cloud thickness ($\Delta R$) equals its distance from the black hole ($R$), the column density ($N_{\rm H}$) and the minimum density ($n_{\rm low}$) are related by
\begin{equation}
N_{\rm H} = n_{\rm low} R.
\end{equation}
Combining this with the definition of the ionization parameter yields 
\begin{equation}
n_{\rm low} =  \frac{\xi N^2_{\rm H}}{L},
\end{equation}
where $\xi$ is the ionization parameter and $L$ is the ionizing luminosity. The maximum distance $R_{\rm upp}$ follows as
\begin{equation}
R_{\rm upp} = \sqrt{\frac{L}{n_{\rm low}\ \xi}}.
\end{equation}
Using the observed parameters, the outflow density exceeds $1 \times 10^{5}$ cm$^{-3}$ during the decay phase and averages around $1 \times 10^{4}$ cm$^{-3}$ in other phases. Due to the loose lower limit on density, the distance constraint remains broad, with $R_{\rm upp} \sim 1\times 10^{5}$ $R_{\rm g}$.   

A second constraint can be obtained by assuming the outflow is just capable of escaping the gravitational potential of the supermassive black hole. The boundary between escaping and failed outflow occurs when the outflow velocity equals $\sqrt{2 G M_{\rm BH} / R}$, where $G$ is the gravitational constant, and $M_{\rm BH}$ = $2.8 \times 10^{7}$ $M_{\odot}$ \citep{bentz2021}. By substituting the escape velocity-based distance into the ionization parameter definition, the density can be expressed as
\begin{equation}
    n_{\rm esc} = \frac{L v^4}{4 G^2 M^2_{\rm BH}\xi},
\end{equation}
where $v$ is the outflow velocity that can be approximated by the radial velocity listed in Table~\ref{tab:Felines} for an order-of-magnitude estimate. As shown in Fig.~\ref{fig:den_dist}, $n_{\rm esc}$ serves as an upper limit, which in turn provides a lower limit on the distance $R_{\rm esc}$. During the decay phase, the outflow density is constrained to be below $10^{12}$~cm$^{-3}$, while the corresponding distance is likely above 50~$R_{\rm g}$.

While the above two approaches offer loose constraints on the density and location of the UFO, we next considered a more speculative scenario to obtain a tighter limit. The UFO is a stream $-$ a broad and relatively diffuse outflow launched vertically from the accretion disk \citep{fukumura2010}. As illustrated in Fig.~\ref{fig:den_dist}, at each phase a portion of this stream passes through our line of sight within the $6.0-10.0$~keV energy range. During the pre-flare, rise, after, and post-flare phases, we observe the stream itself, allowing us to track velocity changes over time. We speculate that a dense and discrete clump, which is embedded within and comoving with the stream \citep{takeuchi2013}, enters our line of sight at the start of the decay phase and exits by its end. The decay phase thus offers a unique $\sim 40$~ks window to directly study the properties of this clump.

Although this is a speculative model, it provides a reasonable explanation for the observed constant column density, which only increases during the decay phase (Fig.~\ref{fig:xi}). An alternative idea is that the clump is always visible but only fully covers the hard X-ray line of sight during the decay phase, with partial covering at other times. While plausible, this alternative scenario would not change the main conclusions derived from our proposed stream model.

Based on the observed radial velocities listed in Table~\ref{tab:Felines}, and an inclination angle of 23$^{\circ}$ reported in \citet{gravity2021}, we can estimate the vertical velocity of the stream at each phase. Assuming that the main clump was launched at the start of the pre-flare phase and followed the stream velocity during its lift-up, its vertical displacement would reach approximately 120~$R_{\rm g}$ by the decay phase. This corresponds to a radial distance of roughly 130~$R_{\rm g}$. Based on the observed luminosity and ionization parameter, the inferred density is approximately 1~$\times$ 10$^{11}$ cm$^{-3}$. This value is consistent with the 
constraints obtained above, $n_{\rm low}$ and $n_{\rm esc}$. It is also in agreement with densities reported in other UFO studies \citep{fukumura2018}.

This geometrical model suggests a launch radius of approximately $\sin(23^\circ) \times 130 = 50~R_{\rm g}$ for the UFO. This location is noteworthy because, according to the standard thin disk model \citep{ss1973}, using the black hole mass of $2.8 \times 10^{7}$ $M_{\odot}$ \citep{bentz2021} and an Eddington accretion rate of 0.07 \citep{2007MNRAS.378..649S}, the dynamical timescale of the disk at 50~$R_{\rm g}$ is estimated to be $T_{\rm dyn} \sim 1 \times$ 10$^{5}$~s.  Interestingly, this timescale closely matches with the observed duration of the soft X-ray flare (Fig.~\ref{fig:lc}). At the same radius, the disk rotation timescale is $ 2\pi T_{\rm dyn} \sim 6 \times$ 10$^{5}$~s, suggesting that orbital motion likely has a limited impact on the outflow kinematics over the observed period.

Our geometrical scenario, as illustrated in Fig.~\ref{fig:den_dist}, further suggests that the clump obscures and passes through the X-ray emission region within the decay phase, over a timescale of no more than $4\times 10^{4}$~s. This implies that the transverse travel distance is at least equal to the combined diameters of the clump and the X-ray corona at 8~keV. Assuming the clump fully covers the emission region during this phase, both the clump and the hard X-ray corona would have diameters of 12~$R_{\rm g}$. If the covering factor is less than 1, the clump size would decrease while the corona size increases. As shown in Fig.~\ref{fig:xi}, the minimal covering factor is about 0.4, corresponding to a clump diameter of 10~$R_{\rm g}$ and corona size of 16~$R_{\rm g}$. Therefore, we considered 12~$R_{\rm g}$ to be the upper limit for the clump diameter.

 Assuming that the outflow stream also has a thickness of 12~$R_{\rm g}$, we derive a lower limit on the density of the clump of approximately $3\times 10^{9}$~ cm$^{-3}$. This places the radial distance of the outflow during the decay phase at less than 1000~$R_{\rm g}$ (Fig.~\ref{fig:den_dist}). These constraints are consistent with all the previous limits derived earlier. This estimate is based on the assumption of a spherical clump; if the clump itself is stream-like with a smaller thickness, the actual radial distance could be further smaller.

In summary, the observed properties can be adequately explained by a relativistic outflow originating at about 50~$R_{\rm g}$ from the innermost region of the disk. This is in broad agreement with the UFO detected in PDS~456 with XRISM \citep{2025Natur.641.1132X}. The prominent feature during the decay phase can be attributed to a major clump within this flow, with a density of about 1~$\times$ 10$^{11}$~cm$^{-3}$.

\subsection{Constraints on the velocity and acceleration}
\label{dis:velocity}
\begin{figure*}[!htb]
    \centering
    \includegraphics[width=1.\linewidth]{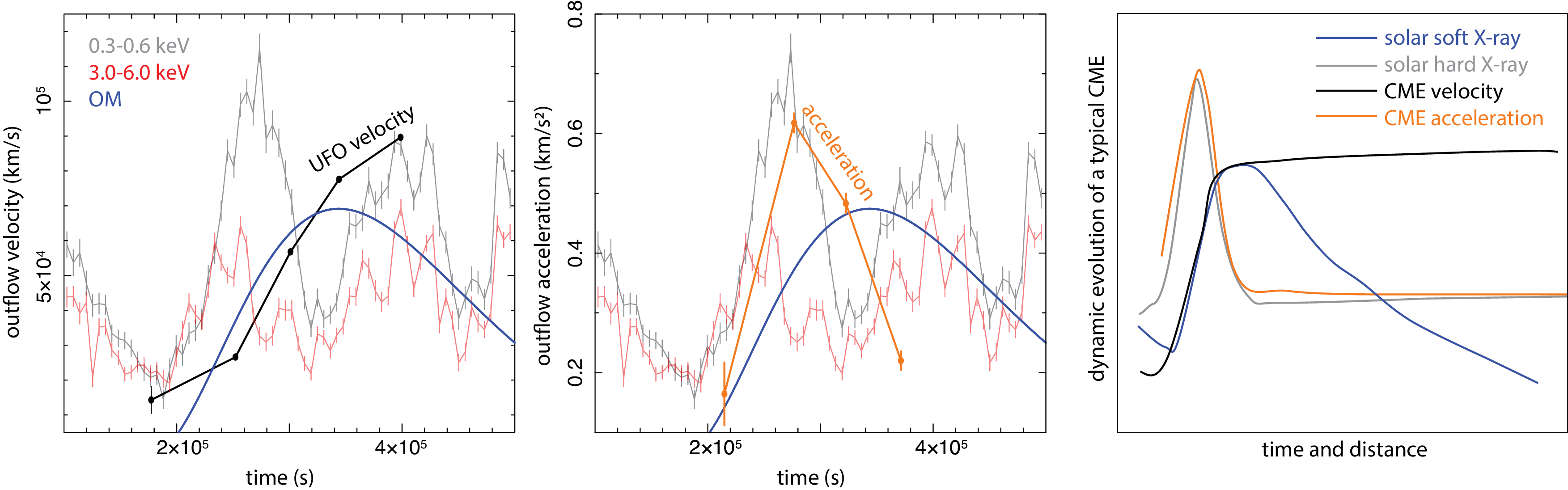}
    \caption{Comparison of UFO and CME dynamics. Left: Evolution of the UFO velocity (in black).  Middle: Evolution of the UFO acceleration (in orange).\ Both panels are overlaid with the \textit{XMM-Newton} OM (blue), soft-band (gray), and hard-band (red) X-ray light curves. Right: Schematic view of the early dynamic evolution of a typical CME event close to the Sun, adapted from \citet{temmer2016}. We see that the UFO and CME patterns are similar to each other, with the acceleration phase aligning with the hard-band flare and the velocity correlating with the soft-band flare.  }
    \label{fig:ufoacc}
\end{figure*}

In Fig.~\ref{fig:ufoacc} we show the evolution of the outflow velocity obtained from the Resolve data, along with the corresponding acceleration. The best-fit velocity increases from 0.05$c$ to 0.3$c$ over about $2 \times 10^{5}$~s, with the acceleration peaking at roughly 0.6~km~s$^{-2}$ around the time of 2.8$\times 10^{5}$~s.

Assuming that the outflow is launched at 50~$R_{\rm g}$, the observed radial velocity of 0.05$c$ at the pre-flare phase (Table~\ref{tab:Felines}) falls well below the escape velocity of 0.19$c$. This implies that the outflow in the pre-flare phase would not escape without significant acceleration. The acceleration that in fact occurs during the soft flare, however, propels the outflow to 130~$R_{\rm g}$ by the decay phase. At this point, its observed velocity of 0.19$c$ (Table~\ref{tab:Felines}) now significantly exceeds the local escape velocity of 0.03$c$, confirming that the outflow has successfully escaped and is propagating away.

The direct acceleration on the wind gained by absorbing or scattering photons can be calculated as
\begin{equation}
a = \frac{\int F(E) [1 - T(E)] dE}{c f m_{\rm p} N_{\rm H}},
\end{equation}
where $F(E)$ is the incoming energy flux, $T(E)$ is the transmission, $c$ is light speed, $m_{\rm p}$ is the proton mass, $N_{\rm H}$ is the column density, and $f$ is the dimensionless factor related to the mass density via $\rho = f n_{\rm H} m_{\rm p}$, with $n_{\rm H}$ being the hydrogen density. Using the observed flux, column density, and ionization parameter during the decay phase (Table~\ref{tab:Felines}), and assuming a wind density of $1 \times 10^{11}$ cm$^{-3}$, the resulting acceleration is on the order of 3~m s$^{-2}$. This is two orders of magnitude smaller than the observed peak acceleration of 600~m~s$^{-2}$ shown in Fig.~\ref{fig:ufoacc}, indicating that radiation pressure alone is unlikely to drive the observed UFO acceleration.

The mass outflow rate can be estimated as 
\begin{equation}
    \dot{M} = 4 \pi b n_{\rm H} m_{\rm p} R^2 v,
\end{equation}
where $b$ is the covering factor of the outflow, $v$ is the velocity at distance $R$. Taking a density of $1 \times 10^{11}$ cm$^{-3}$ and velocity of 0.2$c$ at distance of 50~$R_{\rm g}$, the mass loss rate is 7$b$ solar mass per year. The actual value of $b$ is unconstrained with current data, though it is likely small. The corresponding mechanical power is about $8b \times 10^{45}$ erg s$^{-1}$, which is roughly the Eddington luminosity scaled by $b$.

In fact, it is unlikely that the acceleration is fully continuous. A stable acceleration of 600~m~s$^{-2}$ would cause the central velocity of the absorption line to increase continuously by 24000~km~s$^{-1}$ during the decay phase, far exceeding the observed velocity dispersion of only about 1200~km~s$^{-1}$ (Table~\ref{tab:Felines}). This discrepancy suggests that the acceleration is impulsive rather than steady. Instead of a smooth and persistent increase, the outflow probably accelerates in short and intense bursts, followed by periods of much weaker acceleration. Such impulsive behavior is characteristic of energetic solar phenomena (\citealt{vrsnak2007} and references therein), which will be discussed in the following section.

\subsection{Comparison with coronal mass ejection events}

We further compare the evolution of UFO velocity and acceleration with the X-ray and optical/UV light curves in Fig.~\ref{fig:ufoacc}. The optical/UV light curve is derived from \textit{XMM-Newton} Optical Monitor (OM) data, as outlined in Appendix \ref{app:om}. We observe a 6.6\% increase in the UV flux of NGC~3783 following the soft X-ray flare, with its temporal behavior consistent with a smooth burst model. Interestingly, the UFO velocity appears to correlate with the rise in UV flux, and the peak outflow acceleration coincides with the maximum in the $0.3-0.6$~keV band. The hard X-ray light curve reaches its maximum before both the UV and soft X-ray light curves.

It has been proposed that some of the UFOs might originate from MHD acceleration \citep{fukumura2018}. This scenario may resemble the physics of coronal mass ejections (CMEs), which are massive eruptions of plasma and magnetic fields from stellar coronae into space. CMEs are often observed alongside flares, current sheets, and plasma eruptions, likely driven by rapid magnetic reconnection \citep{dm1998,patel2020}. Statistical analyses of CMEs during their early near-Sun evolution \citep{zhang2006, temmer2016} suggest a two-phase process: an initial slow rise lasting tens of minutes, followed by a rapid acceleration to reach maximum velocity. Comparing the temporal evolution of observed UFO and CMEs reveals potential similarities: the velocity appears to track emission in the longer wavelength (soft X-ray for CMEs and UV for the UFO), while the acceleration correlates better with emission at shorter wavelength (hard X-ray for CMEs and soft X-ray for the UFO).

To compare their characteristic outflow velocities, we utilized the Alfv\'en speed as a proxy,
\begin{equation}
    v_A = \frac{B}{\sqrt{\mu_0 \rho}},
\end{equation}
where $B$ is the field strength, $\mu_0$ is permeability of free space, and $\rho$ is the mass density. For CMEs, adopting $B = 10$~Gauss and $\rho = 10^{-15}$ g cm$^{-3}$ in the corona \citep{asch2005}, the Alfv\'en speed is about 1000~km s$^{-1}$. As for the UFO, assuming a magnetic field strength of 10$^{4}$~Gauss and mass density of $10^{-15}$ g cm$^{-1}$ in the AGN corona \citep{merloni2001}, the Alfv\'en speed becomes about 0.3$c$. This scaling shows a rough agreement with the observed outflow velocities in typical CMEs and the UFO in NGC~3783. The assumed field strength of 10$^{4}$~Gauss is consistent with the maximum expected value derived from the $\alpha$-disk prescription \citep[see][Eq. 2.19]{ss1973}.

Following, for example, \citet{oh1997} and \citet{tsuneta1997}, the total energy release rate from a magnetic reconnection event can be described by
\begin{equation}
\label{eq:energy}
     \frac{B^2}{2 \pi} v_{\rm inflow} L_{\rm r} \geq \frac{d(E_{\rm UFO} + E_{\rm R})}{dt},
\end{equation}
where $v_{\rm inflow}$ is the reconnection inflow speed, $L_{\rm r}$ is the characteristic scale of the reconnection region, $E_{\rm UFO}$ is the kinetic energy of the UFO, and $E_{\rm R}$ is the total radiative energy enhancement in the flare. The left side represents the energy input from reconnection, which is likely larger than the right side, as some energy may be channeled into other forms such as conductive flux.

Here we assumed that the reconnection region has a scale similar to the clump size of the UFO, $L_{\rm r} = 12 R_{\rm g}$ (see details in Sect.~\ref{dis:density}). Taking $B = 10^{4}$ Gauss and $v_{\rm inflow} = 0.1 v_{\rm A}$ \citep{dm1998}, the total energy release rate through reconnection is approximately $2\times 10^{44}$ erg~s$^{-1}$. This is an order-of-magnitude estimate. Comparing this with the observed broad-band X-ray luminosity enhancement during the flare, which is roughly $2\times 10^{43}$ erg~s$^{-1}$, indicates that about 10\% of the injected energy is radiated away in X-ray.

Furthermore, as discussed in Sect.~\ref{dis:velocity}, the kinetic power of the outflow is estimated as $8b \times 10^{45}$ erg~s$^{-1}$, where $b$ is the UFO covering factor. From Eq.~\ref{eq:energy}, this implies a small covering factor of $b \leq 0.02$. The small covering is physically plausible, considering the potential filamentary nature of the outflow suggested in Fig.~\ref{fig:den_dist}.

As reported in, for example, \citet{kontar2017}, the velocity dispersions of the CME ejecta are on the order of $\sim 100$~km s$^{-1}$, which are much lower than their bulk velocities of $\sim 1000$~km s$^{-1}$. This suggests that a coherent magnetic structure driving the CME would suppress small-scale motions. Similarly, as shown in Table~\ref{tab:Felines}, the bulk flow represents the dominant form of kinetic energy in the UFO in NGC~3783.

Evidence of CME-like eruptions during soft X-ray flares has been observed in Galactic microquasars \citep{fender1999} and in $\mathrm{Sgr~A^{*}}$. Although the origin of this correlation remains uncertain, theoretical and simulation studies suggest that rotating magnetic field lines inflate and reconnect above the disk, rapidly heating the corona during flares, while simultaneously expanding and launching plasma-loaded magnetic flux tubes that constitute the UFO \citep{yuan2009}. Reconnection could involve poloidal fields \citep{em2023}, toroidal fields on the disk \citep{machida2003}, or a combination of the two. A few MHD simulations have also explored the connection between such eruptions and various accretion disk configurations, including magnetically arrested disks \citep{narayan2003}.

It is plausible that magnetic reconnection-triggered UFOs induce quasi-periodic X-ray flares \citep{rip2020}. The periodicity of these flares is crucial for constraining the reconnection rate. While the X-ray light curve of NGC~3783, shown in Fig.~\ref{fig:lc}, exhibits features resembling quasi-periodic variations, a comprehensive quantitative search for periodicity remains nontrivial and will be addressed in a subsequent study. The XRISM campaign on NGC~3783 offers a promising vision; to fully understand UFO launching and energy transport in the innermost regions of the accretion disk, it is still essential to collect more flare–UFO cases across different systems. Long-term monitoring and high-resolution spectroscopy with XRISM and  upcoming missions such as {\it NewAthena} \citep{cruise2025, philipe2025} will be vital for advancing this research.

\section{Summary}
\label{sec4}

The ten-day spectroscopic campaign that observed NGC~3783, conducted with XRISM in combination with six other X-ray and UV observatories, offered an unprecedented opportunity for unexpected discoveries. The X-ray light curve shows multiple flares, including a major event characterized by initial hard X-ray peaks followed by a strong soft X-ray peak. During the decay of this soft flare, an absorption feature at 8.4~keV in the Resolve spectrum suggests the presence of a UFO with a radial velocity of 0.19$c$ ($\sim 57000$ km~s$^{-1}$). This outflow may not be an isolated event but rather part of a broader outburst that lasted approximately three days. A secondary outflow, with a velocity of about 3700~km~s$^{-1}$, was also detected coincident with the UFO. Based on the physics constraints derived from the Resolve data, it is plausible that magnetic fields within the accretion disk serve as the central engine driving the UFO launch in this low-$L/L_{\rm Edd}$ AGN. This mechanism may also be linked to the observed soft flare.

\begin{acknowledgements}

SRON is supported financially by NWO, the Netherlands Organization for
Scientific Research. KF acknowledges support from NASA grant 80NSSC23K1021. Part of this work was performed under the auspices of the U.S. Department of Energy by Lawrence Livermore National Laboratory under Contract DE-AC52-07NA27344. The material is based upon work supported by NASA under award number 80GSFC21M0002. EB acknowledges support from NASA grants 80NSSC20K0733, 80NSSC24K1148, and 80NSSC24K1774. M. Mehdipour acknowledges support from NASA grants 80NSSC23K0995 and 80NSSC25K7126. This work was supported by JSPS KAKENHI Grant Numbers JP21K13963, JP24K00672, JP24K00638, JP21K13958, JP24K17104, JP25H00660. AT and the present research are in part supported by the Kagoshima University postdoctoral research program (KU-DREAM). KF is grateful to Emeritus Professor of Kyoto University, Kazunari Shibata, for a fruitful discussion on magnetic reconnection. MS acknowledges support through the European Space Agency (ESA) Research Fellowship Programme in Space Science. M. Mizumoto acknowledges support from Yamada Science Foundation.

\end{acknowledgements}

\bibliographystyle{aa}
\bibliography{main}

\begin{appendix}
\onecolumn

\section{Look elsewhere effect in the phase-resolved spectra}
\label{app:lee}

\begin{figure*}[!htbp]
\centering
    \includegraphics[width=0.95\linewidth]{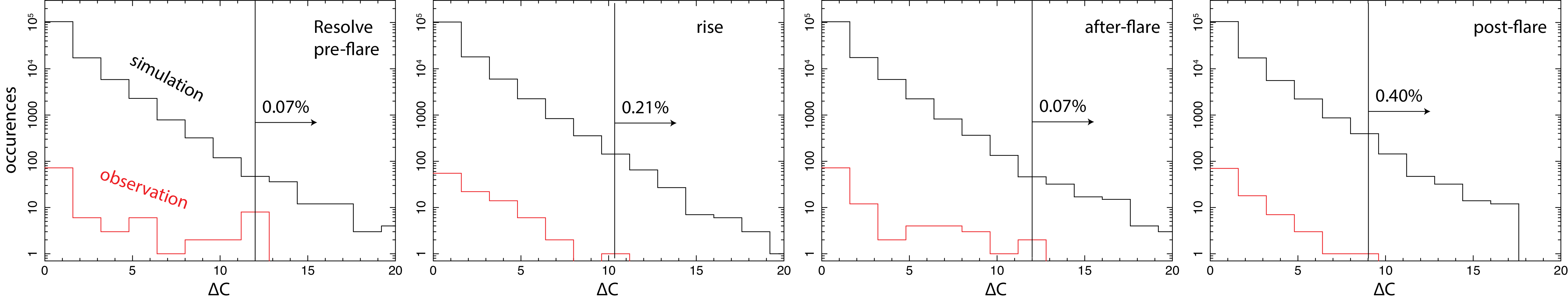}
    \caption{Histogram of $\Delta C$ distributions for Resolve. The black histogram shows the results from $1.3 \times 10^{5}$ Monte Carlo simulations per phase, and the observed distribution is shown in red. Vertical lines mark the positions of the best-fit features.  }
    \label{fig:allphase}
\end{figure*}

In this appendix we present additional information that addresses the LEE in the phase-resolved spectra of the soft flare. The approach has been detailed in Sect.~\ref{decayufo} for the potential detection of UFOs in the decay phase. Here, we applied the same method to the other phases and provide the corresponding significance. 

For each phase, we simulated a set of $1.3 \times 10^{5}$ spectra for Resolve, and $1 \times 10^4$ for other instruments, based on the baseline model. For each simulated spectrum, we scanned the $7.0-9.0$~keV range with the {\tt pion} model for the UFO and computed the $\Delta C$ relative to the baseline. In Fig.~\ref{fig:allphase} we plot the observed and simulated $\Delta$C diagram for Resolve. The original $\Delta$C improvements of 12, 10, 12, and 9 for the pre-, rise, after-, and post- flare phases, become 11.5, 9.5, 11.4, 8.3 after correcting for the LEE. The effect of LEE is reflected in the significance shown in Table~\ref{tab:Felines}. 

For all phases, the data from the CCD instruments agree with the best-fit model shown in Table~\ref{tab:Felines}. In the decay phase, the collection of CCD data boosts the significance of the UFO feature. In the other phases, since the possible features are much weaker, the LEE effect becomes more dominant; therefore, the relative boost from the CCD data to the total significance is less pronounced than in the decay phase.

\begin{figure*}[!htbp]
    \centering
    \includegraphics[width=0.7\linewidth]{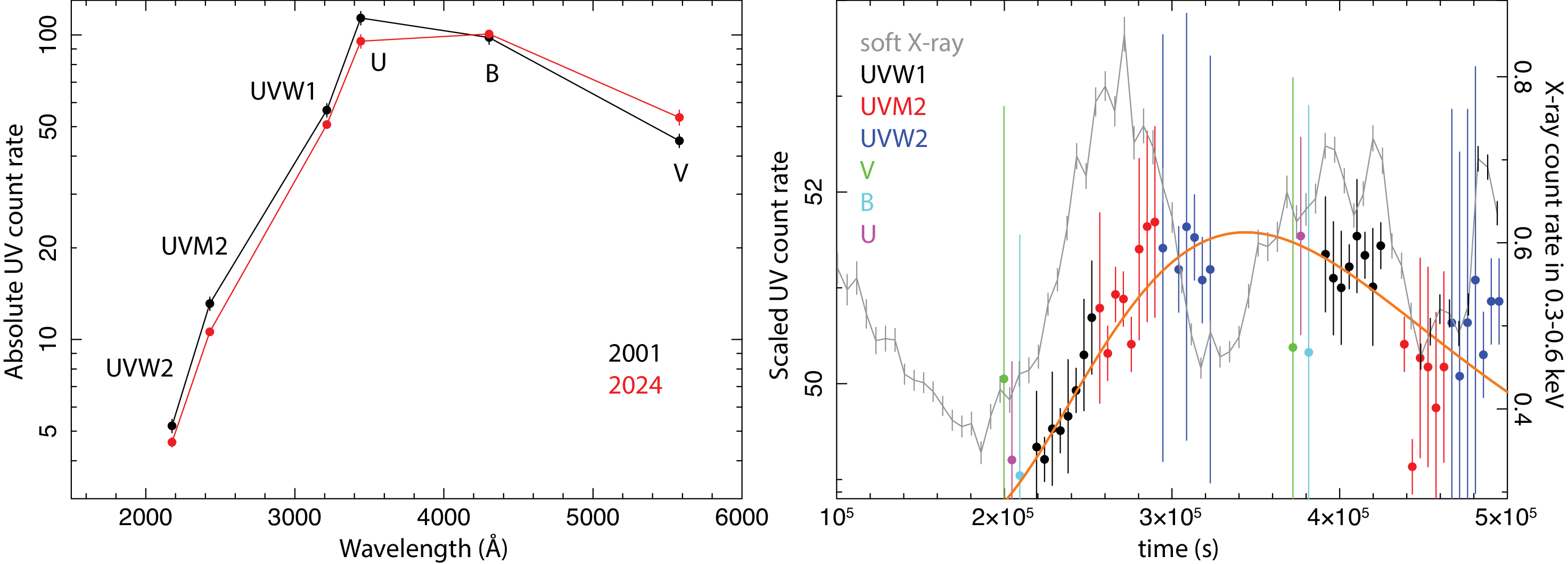}
    \caption{Left: Comparison of the time-averaged \textit{XMM-Newton} OM spectrum from the current 2024 campaign with 2001 data. Right: Fluxes of different OM filters, scaled to the UVW1 filter, derived from the 2024 time-average spectrum. The best-fit phenomenological model is shown in red. The $0.3-0.6$~keV X-ray light curve is overlaid for comparison. }
    \label{fig:xmmom}
\end{figure*}

\section{\textit{XMM-Newton} Optical Monitor data}
\label{app:om}

The \textit{XMM-Newton} OM observed the AGN using a sequence of different filters, with one filter active at a time. As shown in Fig.~\ref{fig:xmmom}, the average UV photometry is broadly consistent with archival UV data from 2000–2001, although the new spectrum appears slightly softer. This likely reflects a change in the disk component.

To construct a continuous light curve from the discontinuous photometric filter data, we calculated the average count rate ratios between UVW1 and the other filters, using the time-averaged data. This enabled us to scale the count rates from other filters to the UVW1 reference. In doing so, we assumed the optical/UV spectral shape remained constant during the observation, while the overall flux varied. As shown in Fig.~\ref{fig:xmmom}, the scaled data form a quasi-continuous light curve, capturing both the soft flare and a subsequent period. The scaled UVW1-equivalent count rate exhibits a smooth rise of approximately 6\%, peaking at $t \sim 3 - 4\times 10^{5}$~s, followed by a gradual decline toward $t \sim 5\times 10^{5}$~s. To model this behavior, we fit the light curve with a smooth burst profile plus a constant component. The fit, shown in Fig.~\ref{fig:xmmom}, describes the light curve well, indicating a peak increase of $6.6 \times 2.1$\% at $3.4 \times 10^{5}$~s. 

According to \citet{mehd2017}, the host galaxy contributes approximately 35\% of the total UVW1 flux. Accounting for this, the observed increase in the OM flux corresponds to about 10\% of the intrinsic UV disk flux. As shown in Fig.~\ref{fig:xmmom}, the OM light curve peaks with a delay of $\sim 1\times 10^{5}$~s relative to the peak of the flare in the $0.3-0.6$~keV.

\end{appendix}

\end{document}